# Exploring AlphaFold 3 for CD47 Antibody-Antigen Binding Affinity: An Unexpected Discovery of Reverse docking

Yiyang Xu[1, #], Ziyou Shen[2, #], Yanqing Lv[3], Shutong Tan[4], Chun Sun[3], Juan Zhang[3, *]

[1] University College London Cancer Institute, 72 Huntley Street, WC1E 6DD, London, United Kingdom
[2] King's College London School of Basic & Medical Biosciences, Great Maze Pond, SE1 1UL, London, United Kingdom
[3] China Pharmaceutical University School of Life Science and Technology，Tong Jia Xiang 24，Nanjing, Jiangsu Province, 210009，China
[4] China Pharmaceutical University School of Science，639 Longmian Avenue, Nanjing，Jiangsu Province, 211198, China
# These authors contributed equally to this work
＊ This author is corresponding author, please contact by zhangjuan@cpu.edu.cn

## Introduction

AlphaFold 3 (AF3) is the latest deep-learning based AI model for predicting a wide range of biomolecular structures, including nucleic acids, proteins, ions and their complexes. The two greatest revolutions of model architecture are 1) Evoformer module and 2) Diffusion module, replacing Pairformer and Structure modules in the AF2. Another achievement of AF3 is the increased predicting capacity by decreasing the multiple sequence alignment (MSA) dependence, which is fulfilled through enhanced calculation efficiency due to application of the Evoformer module [1-2]. An overall-improved complex prediction accuracy is AF3's strongest competitiveness over its predecessor, the AlphaFold-Multimer (AFM) v.2.3. Tremendous improvements in mean accuracy score were obtained by using AF3 compared to AFM for all protein-protein complexes from 67.5 to 76.6, and especially protein-antibody complexes (29.6 to 62.9) [1,3]. Current concern of AF3 is hallucination, attributed to generative diffusion-based module change, generating specious α-helix structures in the disordered regions that should be extended loops [1]. Previous case study also illustrates AF3's inability in structural details' accuracy facing specific protein-ligand complex predictions [4].

  Nevertheless, structural accuracy is not the end. Adequately precise predictions will guarantee more accurate dynamic calculations [5]. Possible applications to predict antibody mutation-induced binding free energy changes have been tested [6-7]. Another potential application of AF3's prediction untested is to compare affinities of antibodies to targeted antigen, which helps novel antibodies' screening. There have been studies attempting to predict affinity of protein-ligand complexes [8], but the higher complexity and fewer PDB reference of protein-antibody complexes' structures makes them require independent validation, posing greater challenges for AF3. Therefore, AF3's effectiveness on ranking affinity for four different CD47-targeting antibodies (all with assessed $K_d$) was tested, simulating real screening. Here, relative binding free energy (RBFE) would be an appropriate indicator of affinity. In those scenarios, AF3's reliability compared to available competitors was also evaluated. During this process, a novel fault pattern was identified as 'reverse docking, which would be caused by the AF3's architecture revolution [1].

  Regarding CD47, it is a relatively new target for cancer immunotherapy, being widely overexpressed on cancer cells in solid and haematological malignancies [9-10]. The common drug mechanism is disrupting their 'Don't eat me' signal by competitive binding of anti-CD47 antibodies to CD47 [9,12]. This prevents CD47 binding with signal regulatory protein-α (SIRPα) on macrophages, thus restoring macrophage phagocytosis [9]. So far, relevant drug development has made significant progress, with a first-in-class anti-CD47 monoclonal antibody Magrolimab (5F9) in Phase III and C47B222 (5TZ2) recorded in the Protein Data Bank (PDB) database in Phase II [11-12]. Unfortunately, there are still no anti-CD47 antibody-based drugs approved due to safety issues [13]. Despite challenges, the CD47 target still holds potential for pharmaceutical innovation.

## Methodology

### Sequence Acquisition for Antigen and Antibody Subjects
7XJF, 7WN8, 5IWL (5F9 PDB ID), 5TZ2 and 5TZU antibody sequences were obtained from PDB [13-14]. All the other antibody sequences were laboratory sourced. CD47-extracellular domain (ECD) sequence was taken from Uniprot.

### Molecular Modelling for Antigen and Antibody Subjects
The CD47-ECD molecular model was constructed by the Swiss-Model server. For the lab antibody subjects D0604, D2510, 5F9 and PDB subject 5TZ2, the molecular modelling was done in Discovery Studio using the antibody homologous modelling method.

### Global Cα RMSD Calculation
The experimental style CD47 antibody-antigen complex structures from PDB were constructed in PyMOL and cleaned by the command "remove not polymer. protein". All the molecular viewings were conducted in PyMOL. In terms of the global Cα RMSD calculation and overlapping graph plotting for the CD47 antibody-antigen complexes of PDB and experimental structures, the built-in alignment tool and contact finding tool in PyMOL were utilized respectively.

### Antibody-Antigen Complex Prediction and Molecular Docking
AF3 and HDOCK were used to predict CD47 antibody-antigen complex structures with sequence of CD47-ECD, where antibody Fab heavy and light chains were entered separately. AFM was also used to predict CD47 antibody-antigen complex structures with sequence of CD47-ECD, where antibody Fab heavy and light chains were entered together. Molecular docking was performed by the ZDOCK and RDOCK modules in Discovery Studio, as well as the PIPER module in Schrodinger. Details about rounds and optimised conformational selection are stated in supplementary material M.8.

### Principle of Molecular Mechanics/Generalised Born Surface Area (MM/GBSA) Calculation [15]
The MM/GBSA calculation was conducted by the Prime module in Schrodinger. The MM/GBSA method is a computational approach that calculates RBFE. In essence, this calculates the difference between the binding free energies of two solvated molecules in the bound and unbound states, or the free energies of different solvated conformations of the same molecule. The fundamental equation is presented below (Eq2).

$$\Delta G^0_{bind} = \Delta G^0_{complex} - (\Delta G^0_{receptor} + \Delta G^0_{ligand})$$

Eq1. Basis for free energy of binding calculations.

In order to avoid the fluctuation of the total free energy exceeding binding energy, the MM/GBSA algorithm splits the total BFE in the solvent into the molecular mechanics term (ΔG$^0$$_{bind, vacuum}$) and the solvation energy (ΔG$^0$$_{slov, complex}$), which are calculated separately as follows (Eq3).

$$\Delta G^0_{bind,solv} = \Delta G^0_{bind,vacuum} + \Delta G^0_{solv,complex} - (\Delta G^0_{solv,ligand} + \Delta G^0_{solv,receptor})$$

Eq2. ΔG$^0$$_{solv, ligand}$, ΔG$^0$$_{solv}$: Free energies of the ligand and receptor monomers in the solvent.

**Design of RBFE Test, Ranking Test and Scoring System**

The 5TZ2 subject with determined experimental structure was chosen. MM/GBSA was first used to calculate RBFE of the PDB experimental structure (RBFE1). All the candidates were asked to predict the 5TZ2 subjects for optimal conformations. Subsequently, MM/GBSA was used to calculate RBFE of the optimal conformations (RBFE2), and the difference between RBFE2 and RBFE1 was calculated as ΔRBFE. Smaller ΔRBFE value represents lower deviation, which were ranked by the scoring system: first 700 points, second 560 points, third 420 points, fourth 280 points and fifth 140 points. The $K_d$ values for 5TZ2/5F9 subjects were obtained from research papers [13,16], while those for D0604/D2510 were obtained from the Fortebio test. The $K_d$ was utilised as an indicator of the true binding affinity ranking. Similarly, all candidates were asked to give predictions for all subjects, and MM/GBSA was used to calculate the RBFE.

    The overall accuracy for each candidate was analysed by the Spearman's rank correlation coefficient. Furtherly, the accuracy of affinity relationship predictions (local accuracy) between antibody pairs was assessed using another scoring system: total score 700 points, with 100 points awarded for each successful prediction of a subject and 50 points for each successful prediction of in-paired affinity relationships. Instead, each affinity relationship being predicted wrongly within one place differs from the fact, for example antibody A-second place and B-third place being predicted to be second and fourth, scores 30 points. A wrong prediction within two places (second/third to first/fourth) scores 10 points, and the opposite prediction is not scored. The final score was generated by weighted calculation over the 2 tests in: Final Score = RBFE Score*0.2 + Ranking Score*0.8.

**Statistical Analysis for the Reverse docking**

To determine whether the 'reverse docking phenomenon for the D2510 subject was a coincidence, AF3 was asked to perform 75 additional sets of predictions. In total, 81 predictions were classified: R = reasonable model, Abo/Ago = antibody/antigen opposite (reverse docking). The significance of the result was calculated as p value. Similar attempts were conducted by AF3 for other D25 antibody family members, including D2512, D2523, D2525 and D2546, for statistical analysis of the 'reverse docking phenomenon, using F and t-test. The benchmark for successful correction was 10 predictions without 'reverse docking.

**Result**

**Experiment 1. Overall structural accuracy evaluation of AF3 in PDB subject predictions**

To confirm the high performance of AF3 in predicting antibody-antigen complexes, particularly CD47 targeting ones, five subjects were selected from the PDB database for analysis. The PDB entries for these subjects are 7XJF, 7WN8, 5F9, 5TZ2 and 5TZU. All their experimental structures are obtained by X-ray diffraction method experimentally. The CD47-ECD was used as the target for these antibodies which represents the main antibody binding domain.

    The primary evaluation index is the global Cα root mean square deviation (RMSD), which reflects the spatial proximity between experimental and AF3 predicted structures. The smaller the RMSD value between two structures, the higher similarity they have. In general, a good score can be RMSD < 3Å [17]. RMSD values are calculated by the alignment tool in PyMOL (see Table 1). All subjects show relatively credible results around the standard level for a good score, except 5F9 that may be due to its complicated diabody structure. However, only 7WN8 and 5TZ2 have succeeded in a stricter scenario, with RMSD values of 1.16Å and 0.79Å respectively. This shows that although AF3 has made great progress, it still struggles in some antibody-antigen complex predictions due to their more complex and massive structures. In the aspect of the evaluation indices of AF3 itself, predicted template modelling (pTM) and interface predicted template modelling (ipTM) scores, they

measure the accuracy of the entire predicted structure [1,18]. To specify, a pTM score above 0.5 indicates a high similarity between the overall predicted and experimental structure. On the other hand, an ipTM score above 0.8 represents a confident high-quality prediction in subunit relative positions within the complex. However, there is a grey zone between 0.6 and 0.8, which means the score within this range could result in an accurate or inaccurate structure, and an ipTM score below 0.6 indicates an unreliable predicted structure [1,19]. As the binding domain for antibody-antigen complex is essential, which simply shows the amino acid residues involved at a molecular level, ipTM score can be effectively applied to show the degree of precision for the predicted structure. The result for the pTM and ipTM scores shows a similar situation, which corresponds to the result of RMSD evaluation. Most subjects have high scores for both indices, where 5TZ2 and 5TZU have the highest pTM and ipTM scores around 0.9, followed by 7XJF. AF3 fails to construct a reliable structure for 5F9 with pTM 0.46 and ipTM 0.36. Interestingly, ipTM score for 7WN8 falls into the grey zone, which is 0.68. In conclusion, AF3 possesses the capability to predict CD47 antibody-antigen complexes with a relatively high accuracy.

**Table 1. Comparison of Global Cα RMSD for the CD47 Antibody-Antigen Complexes Between AlphaFold 3 Model and X-Ray Diffraction Structures with AlphaFold 3 ipTM and pTM Scores.**

| PDB | Antibody | Type | Resolution (Å) | AlphaFold3 Global Cα RMSD (Å) | ipTM | pTM |
|---|---|---|---|---|---|---|
| 7XJF | 6MW3211 | Fab | 2.6 | 3.73 | 0.86 | 0.88 |
| 7WN8 | BC31M5 | Fab | 2.8 | 1.16 | 0.68 | 0.77 |
| 5IWL | 5F9 | Diabody* | 2.8 | 9.08 | 0.36 | 0.46 |
| 5TZ2 | C47B222 | Fab | 2.3 | 0.79 | 0.89 | 0.90 |
| 5TZU | B6H12.2 | Fab | 2.1 | 5.96 | 0.91 | 0.91 |

*Diabody of Hu5F9-G4 refers to fusion of the heavy and light variable domains (VH and VL) with a short GGSGG linker for Hu5F9-G4 diabody/CD47-ECD crystallization that presents as a symmetric, domain-swapped dimer linking 2 copies of the CD47-ECD. Since crystallization of neither fragment antigen-binding (Fab) nor single-chain variable fragment (scFv) formats are successful [14].

To further explore the specific interacting amino acid residues within the binding domain, 5TZ2 is selected as an example to perform overlapping alignment analysis using the same alignment tool in PyMOL (see Fig. 1). The enlarged views of the interacting residues are shown on the left in Fig. 1a and 1b for the experimental and predicted structures, where any contacts shown between C47B222 (5TZ2 antibody) and CD47 are restricted within 3Å. Fig. 1c displays the overlapping view of the experimental and predicted structures, which shows that the binding domain emerges mainly between the Fab heavy chain and CD47. The overall degree of structural overlap is high, which also confirms the high RMSD value for 5TZ2. The entire sequence for the antibody C47B222 is displayed in the lower panel, with heavy chain (H) and light chain (L). Clearly, the antibody residues involved in the binding domain are ASP-55, ARG-59 and HIS-105 for the experimental structure, which all emerge in the predicted structure. Even though two extra residues ARG-32 (L) and ARG-101 are predicted by AF3, these may be filtered by the comparative testing over different predictions. Therefore, AF3 has the potential for successfully predicting different antibody-antigen complex structures and even a wider range of biomolecules with high accuracy and precision, which is quite useful for the subsequent analyses of specific physical or chemical properties, whereas challenges still exist.

In summary, AF3 can perform relatively highly accurate predictions for CD47 antibody-antigen complexes, confirmed by RMSD evaluation. The structural precision is extremely high for subject 5TZ2 by seeking the residues involved in the binding domain from comparative analysis, which shows that the experimental and predicted structures are highly overlapped. On the basis of sequence information of the C47B222 antibody, all three residues ASP-55, ARG-59, HIS-105 on the Fab heavy chain presented in the experimental structure for binding are successfully predicted by AF3. This lays a solid foundation for the latter experiments of RBFE and ranking tests in terms of antibody binding affinity comparison at the molecular energy level, as well as further applications in the antibody engineering and pharmaceutical industry.

**Figure 1. Comparative Analysis of Experimental and AF3 Predicted Structures for C47B222 (PDB: 5TZ2)-CD47 Antibody-Antigen Complexes.**

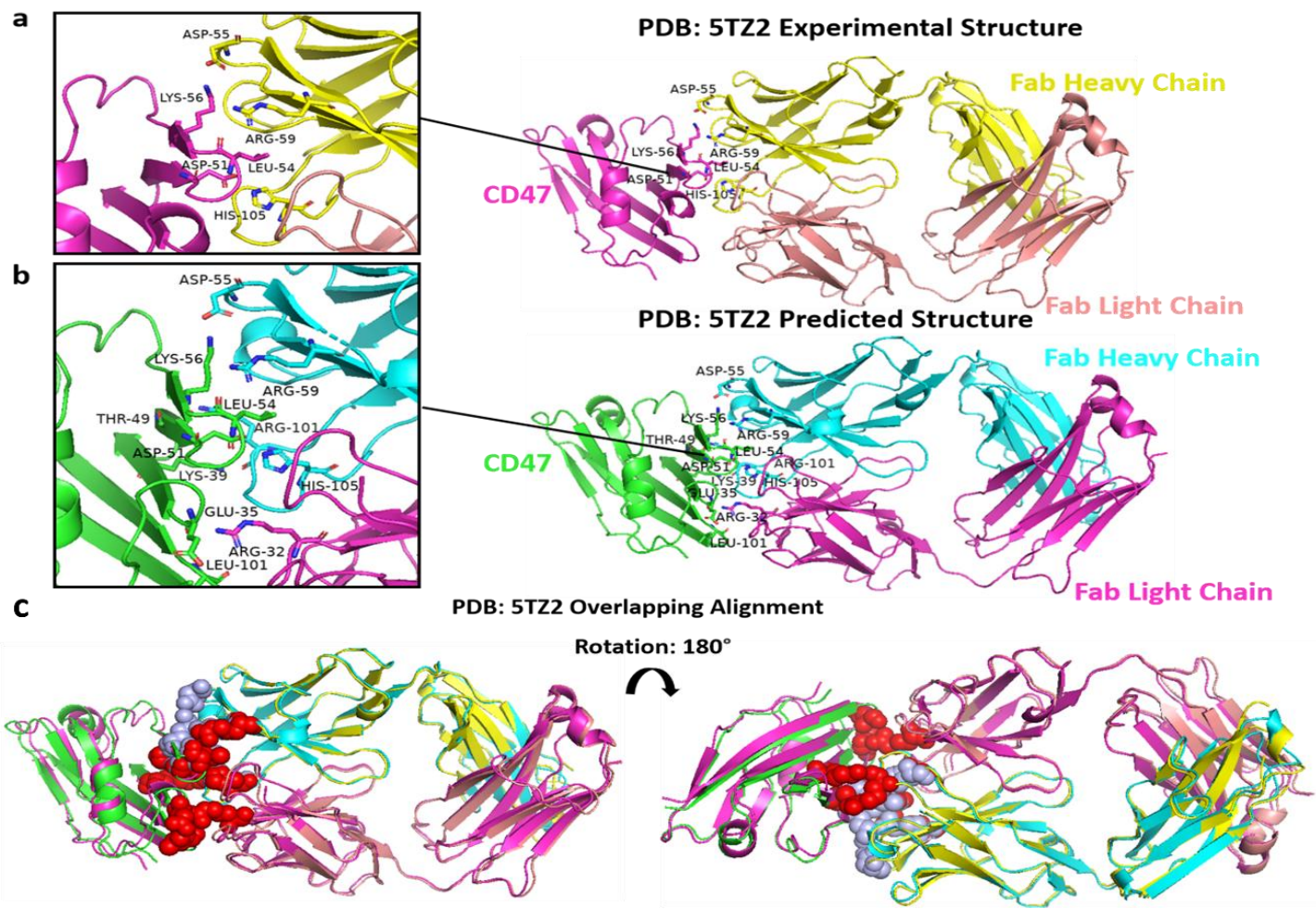

a The experimental structure of the 5TZ2 antibody-antigen complex. Antigen CD47 is shown in pink and antibody is shown in yellow (Fab heavy chain) and salmon (Fab light chain). All the amino acid residues involved in binding domains between antigen and antibody within 3.0 Å are shown by labels, with ASP-51, LEU-54, LYS-56 for antigen and ASP-55, ARG-59, HIS-105 for antibody. b The predicted structure of the 5TZ2 antibody-antigen complex. Antigen CD47 is shown in green and antibody is shown in cyan (Fab heavy chain) and pink (Fab light chain). All the amino acid residues involved in binding domains between antigen and antibody within 3.0 Å are shown by labels, with GLU-35, LYS-39, THR-49, ASP-51, LEU-54, LYS-56, LEU-101 for antigen and ARG-32 (L), ASP-55, ARG-59, ARG-101, HIS-105 for antibody. Note that ARG-32 here is the only residue on the Fab light chain of the antibody in the predicted structure. c The overlapping alignment of experimental and predicted structures in (a) and (b) using the same colour. The two viewing planes are generated by 180° rotation. The overlapping degree for the binding domains of the two structures is shown in bubble clusters with experimental structure in watchet blue and predicted structure in red. The panel at the bottom shows the complete amino acid sequence of the antibody with heavy chain above and light chain below. The residues involved in the binding domain for the predicted structure are indicated in red, and those for the experimental structure are highlighted in grey. The common residues are ASP-55, ARG-59 and HIS-105.

**Experiment 2.1 RBFE test: AF3 prediction accuracy assessment compared to other candidates**

To ascertain whether AF3 has an advantage in terms of accurate RBFE prediction. All structures' RBFE was calculated by MMGBSA algorithm using relevant module on Schrödinger to minimise computational error. Subject 5TZ2 with determined 3D structure was selected, where its RBFE1 is -90.91 kJ/mol [13]. In comparison, AF3 gave the closest RBFE2 of -79.7 kJ/mol (ΔRBFE = 11.21 kJ/mol) scored 700 points, followed by AFM (RBFE2 = -108.37 kJ/mol, ΔRBFE = 16.3 kJ/mol), and the commercial docking candidates (ZDOCK/PIPER) ranked fourth and fifth respectively (see Table 2 and Fig. 2). The advantages of AF3 and AFM are likely based on the complete knowledge of the structure, allowing the MSA module to find high-quality information and, in this case, possibly giving it more reliable interface prediction capabilities for homology modelling than free docking. However, in practical applications, it is impossible to predict only for antibody-antigen complexes with known structures or those with known approximate structures. As with the challenges encountered in the development of CD47 antibodies, laboratories generally have a large number of newly discovered candidate antibodies that require determination of their affinity and other properties [9,11]. Due to the considerable technical difficulty and cost of structural determination (Cryo-Electron Microscopy or X-ray Crystallography), only a few individuals that demonstrate significant clinical potential in subsequent experiments have clear structural information [12-13]. It is therefore reasonable to hypothesise that the ability to at least clearly compare the performance differences between candidate antibodies and existing measured competitors will be one of the most significant forms of progress.

**Table 2. RBFE2 and ΔRBFE for the Optimal Conformation with Candidate Ranking and Score for the RBFE Test (PDB RBFE1 = -90.91 kJ/mol).**

|  | AF3 | AFM | HDOCK | ZDOCK | PIPER |
|---|---|---|---|---|---|
| RBFE2 (kJ/mol) | -79.70 | -108.37 | -40.36 | -23.98 | -16.20 |
| ΔRBFE (kJ/mol) | 11.21 | 17.46 | 50.55 | 66.93 | 74.71 |
| Rank | 1 | 2 | 3 | 4 | 5 |
| Scores | 700 | 560 | 420 | 280 | 140 |

**Experiment 2.2 Ranking test: AF3 prediction accuracy assessment compared to other candidates**

The accuracy of AF3's structural predictions at energy level must be considered with practical application. A meaningful example is to compare novel antibodies with published competitors in terms of antibody-antigen binding affinity. In this study, all candidates underwent the affinity rankings for four antibody subjects (D0604, 5TZ2, D2510, 5F9), with assistance of MM/GBSA calculation.

In the process, only AF3, ZDOCK and PIPER were capable of predicting all complexes (see Table 3). The correlation coefficient between the predicted and true ranking for AF3 is 0.0, indicating its inability to give effective ranking. The correlation coefficients for ZDOCK and PIPER are -0.2, suggesting that the rankings are contrary to the facts. The ranking trends predicted by AF3, ZDOCK and PIPER, as illustrated in line charts, demonstrate discrepancies that are consistent with the conclusions drawn from the correlation coefficients when juxtaposed with the fact ranking trends. For complex affinity sorting of multiple antibodies, all of the current candidates were unable to be used in realistic applications. The local accuracy was subsequently scored, ZDOCK and PIPER received the highest scores (550), followed by AF3 (470). Consequently, ZDOCK, AF3 and PIPER were shown to accurately predict in-pair relationships between antibody affinities. Overall, AF3 showed high-level accuracy for antibody-antigen complex prediction in line with other commercial docking candidates

(ZDOCK/PIPER), in terms of summary score (AF3-516, ZDOCK-496, PIPER-468) (see Fig. 2 and Table 4). The advantage of AF3 over AFM in the prediction of complex antibody-antigen interaction was found (AFM-296 points, difference: 220 points) and believed to be attributed to architectural optimisation.

To specify, architecture revolutions may benefit AF3 in two principal ways. Firstly, MSA processing is reduced from 48 to 4 blocks in AF3, with the outer product combining the MSA into the pairwise representation, which simplifies the complexity of computation. Secondly, the Diffusion module replaces the Structure module, where the latter relies solely on MSA and pairwise representation to assemble protein structure based on relative positions of tokens, that is similar to a person assembling LEGO blocks, but only with information for 'some of the blocks should be interlocked'. In contrast, the Diffusion module is 'Text to Image', wherein the pair-/single-representations are processed as 'condition' inputs, guiding the denoising of initial random noise, generating entire structures in a directional manner. The module is then directed to optimise the overall/local prediction by iteratively adding and removing noise at different noise levels [1]. These improvements reduce workload per prediction thus increasing computational capacity and allowing AF3 to predict more complex interactions. For AFM, heavy/light chain sequences had to be fused and inputted as a dimeric interaction (heavy + light chain, antigen) due to computational limits [2]. Thus, AFM is unable to understand heavy/light chain interactions on novel D2510/D0604 antibodies, resulting in exaggerated errors, whereas AF3 supports three or more chain inputs and successfully completes prediction of all antibodies (see Fig. 2).

**Figure 2. RBFE Test and Ranking Test results with Additive Scores for the Software.**

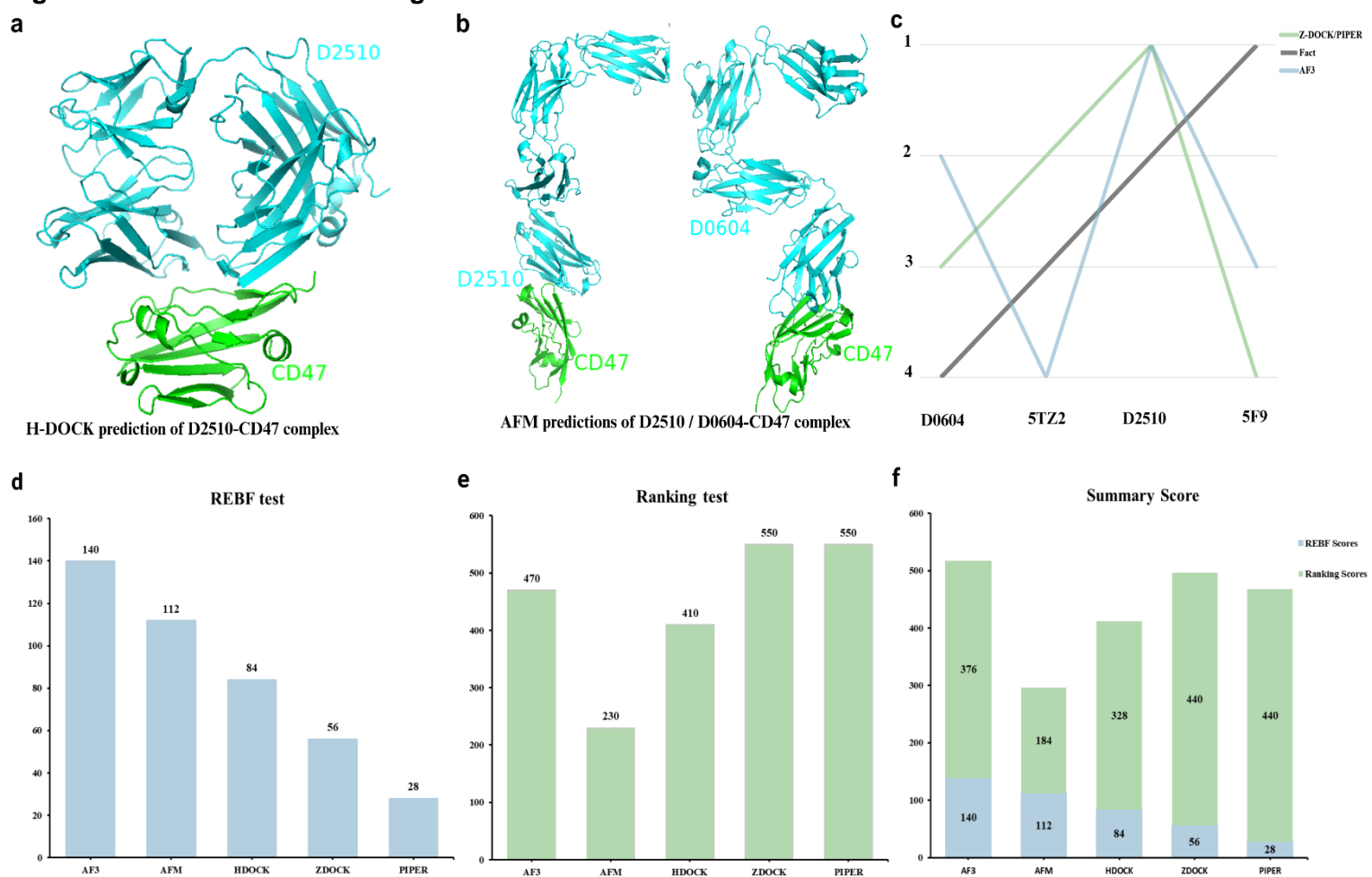

a D2510-CD47 model predicted by HDOCK, with normal D2510 structure visible. b D2510-CD47 model predicted by AFM, with disordered D2510 structure visible, and D0604-CD47 model predicted by AFM, with disordered D0604 structure visible. c Represented by the line graph, the true ranking (grey) is juxtaposed with the predicted ranking for each candidate capable to predict all subjects. d

Scores of all candidates in the RBFE test. **e** Scores of all candidates in the Ranking test. **f** Summary scores of all candidates in weighted components.

**Table 3. Comparison Between Real-$K_d$ and RBFE of Antibody Subjects from Different Candidates (N/A for unsuccessful predictions).**

| Antibody | Real_$K_d$ (M) | ZDOCK RBFE (kJ/mol) | PIPER RBFE (kJ/mol) | AF3 RBFE (kJ/mol) | HDOCK RBFE (kJ/mol) | AFM RBFE (kJ/mol) |
|---|---|---|---|---|---|---|
| D0604 | $8.40*10^{-10}$ | -42.89 | -17.44 | -52.11 | -76.15 | N/A |
| 5TZ2 | $1.12*10^{-9}$ | -23.98 | -16.20 | -79.70 | -40.36 | -108.37 |
| D2510 | $1.62*10^{-9}$ | -23.79 | -12.06 | -29.87 | N/A | N/A |
| 5F9 | $8.00*10^{-9}$ | -106.52 | -114.61 | -55.11 | -46.15 | -61.60 |

**Table 4. Comparison Between Real Antibody Affinity Rankings (Fact) and Ranking Predictions from Different Candidates (N/A for Unsuccessful Predictions).**

| Antibody | Fact | AF3 | ZDOCK | PIPER | AFM | HDOCK |
|---|---|---|---|---|---|---|
| Rk_D0604 | 1 | 3 | 2 | 2 | N/A | 1 |
| Rk_5TZ2 | 2 | 1 | 3 | 3 | 1 | 2 |
| Rk_D2510 | 3 | 4 | 4 | 4 | N/A | N/A |
| Rk_5F9 | 4 | 2 | 1 | 1 | 2 | 3 |

## Experiment 3. Reverse Docking Phenomenon

In the present study, AF3 was found to provide higher confidence scores to a number of incorrect predictions of D2510-CD47 complex that docked the antibody/antigen in an opposite direction to the actual binding site (see supplement Table 5 and Fig. 3) [16]. This suspicious novel fault pattern was named 'reverse docking. In 81 sets, 'reverse docking (Group Ago/Abo) appeared 53 times and all were accompanied by high ipTM+pTM (once ipTM below 0.65; once pTM below 0.74). Reasonable conformations (Group R) appeared 23 times.

The highest ipTM in D2510 relevant 'reverse docking result is 0.62 and highest pTM is 0.71. t-test then determined significantly difference between two groups' means, with a one-tailed p value of 3.83E-33, proving that 'reverse docking is not an occasional phenomenon. The same error was reproduced on the D2523-CD47 complex（see Fig. 3 and supplement Table 6-10).

The cause of this error was suspected to be due to AF3's architecture revolution, mainly in links with reduced MSA dependence. MSA is based on the 'co-evolution' theory, stating a single residue mutation of a protein can only be stably retained through evolution if the corresponding residue in interacting proteins mutate with it. By relying on large-scale protein sequence alignments, MSA has become a widely used strategy to determine the relative spatial relationships of residues [20-23]. In fact, the main reason for MSA dependence reduction in AF3 is the difficulty in finding homologous MSA sequences. Another study that attempted to refine limited MSA demonstrated large MSA sampling was needed in some cases [1, 20, 24-25]. Thus, the MSA module is caught in a dilemma. It is valuable, but the attempt to obtain useful MSA information in protein complex predictions requires an unacceptable efficiency cost. However, the Diffusion and Confidence modules still require MSA information, and reverse docking demonstrates the possibility of simultaneous prediction problems with both modules in AF3: D25 family antibodies do not have access to the reference PDB structure, which may make AF3 more template-dependent. At this point, since the MSA module is down-weighted, it is less likely to provide sufficient information for error correction, as MSA is proved to be intensively dependent on sampling depth and quality [21-22].

Optimisation at the architecture level is complex, and simply increasing MSA blocks is undoubtedly a degradation and the quest for efficiency. A solution was then proposed at the input level: by completing the input sequence of the CD47 transmembrane region (5*α helix), the docking direction of D2510 to CD47 can be efficiently normalized. Totally 20 predictions for D2510 - Full CD47 structure (2 sets of benchmarking test) were made to exclude possibility of coincidence for normalization,19 of the results have ipTM value = 0.43 and pTM = 0.55, and one exception having pTM = 0.56, which are much lower than scores for previous 81 sets. This suggests that AF3 may understand the nature of interactions of common protein structures, and thus the provision of additional reference structures may help to correct fault patterns, but the longer amino acid sequence with more complex structure makes reliable predictions more difficult to get. However, for the D2523-CD47 complex, the same method was ineffective (D2523 bound to alpha helix), and the reason for this instability has not yet been specified (see Fig. 4).

## Discussion

### Insufficiencies of the Experimental Protocol

In this study, deleting CD47 unresolved regions from PDB reference may change the protein conformation. Also, since post-translational modifications (PTMs) such as glycosylation of antigen may affect its interaction with antibody, which are now available for specific residues in AF3 input [1，23], such ignored modifications on CD47 may affect predictions.

In the future, as open-sourced AF3 application attempts, its performance may change due to different types of biomolecules, various cumulative number of times for prediction and more or less resolved structures available in the database. With regard to binding free energy, the strengths and weaknesses of AF3 in predicting antibody-antigen complex interactions were examined in comparison to other competitors, particularly with its predecessor AFM. However, AF3 was declared to be more powerful in RNA-protein, peptide-protein, and small molecule-protein complex interactions compared to its previous version [1, 24]. It is hypothesized that further experimentation, employing a methodology analogous to that employed in this study, would yield more practical applications.

In addition, it is recommended to address the particular finding of the reverse docking phenomenon. However, it has been observed that introducing more reference structures is often accompanied by lower 'ipTM+pTM' scores (see Fig. 3-4). Furthermore, the addition of more restrictive information from the inputs to avoid similar phenomena does not address the root cause of the problem. Further research should consider this issue from a programming perspective: Inspired by HDOCK, in future optimization inputs are supposed to allow human-suggested information, such as operator directly indicating highly suspicious/certain interaction interfaces which may help AF3 to avoid 'reverse docking; In order to better combine confidence scores with results, follow-up researchers may rethink the importance of the MSA module and the balance between effective sampling and prediction efficiency, or try new training techniques may better match the confidence scores and results.

It is also not yet clear whether this problem will arise when AF3 makes predictions against other types of ligand-protein complex interactions, and further experiments or serendipitous discoveries are needed. The question of whether a new framework can be proposed to solve the problem completely, or whether AF3 can be trained specifically for antibody-antigen complex prediction to improve the situation, is a topic that is likely to be of great interest.

### Future prospects

This study indicates that AF3 demonstrates significant improvements over its predecessor and maintains industry-leading performance by leveraging more advanced deep learning methods and avoiding massive reliance on MSA. Nevertheless, the MSA module remains pivotal. For developers, the emergence of 'reverse docking' suggests that efforts to enhance model performance through MSA module optimisation, particularly improving the quality of MSA information input, are highly worthwhile [21]. Indeed, a predictive model utilising artificial intelligence for MSA ranking has already demonstrated superior performance compared to other highly MSA-dependent competitors, especially in the case of antibody-antigen complexes and other protein quaternary structures [25].

Nevertheless, the implementation of advanced deep learning techniques in structure prediction models should continue. Given that AF3 exhibits relatively frequent hallucination when processing complex and long amino acid sequences (see Fig. 4), leveraging deep learning to acquire restraint information might better constrain the emergence of overly free structures. An intriguing concept is to integrate strategies from physics-based docking software with prediction models to normalize structure generation [26]. Further exploration of this is eagerly anticipated, as no novel strategy has been formally announced that has the potential to substantially surpass AF3 in its capacity to predict protein-protein complex structures.

The accuracy and flexibility of the structure generated by AF3 are the basis. This can be evidenced by another study, which combined the deep learning tool AF with a physics-based replica exchange docking algorithm. Its performance surpassed AFM, especially on antibody-antigen complexes [27]. Additionally, it should be noted that AF3 has a faster single task completion time than other models tested, with 5 minutes in average. This makes it more responsive than other models to the high-throughput, high-efficiency demands of real-world antibody drug screening. In the future, if the overall accuracy of affinity comparison is furtherly improved, the affinity performance of new antibodies can be directly predicted by AF3 in comparison with known antibody competitors in the market, so as to quickly screen subjects for pharmaceutical potential, followed by cell cultivation and wet experiment validation [28]. Furthermore, it may even be possible to rapidly screen potential antibodies by computational random mutation of antibody sequences [6-7], achieved by AF3 high-throughput comparison, replacing traditional screening experiments to obtain new antibodies, e.g. hybridoma technology, phage display technology.

**Figure 3. Examples of 'Reverse docking and 'Rational Conformation' for the AF3 prediction of the D2510-CD47 model, and the D2523-CD47 model.**

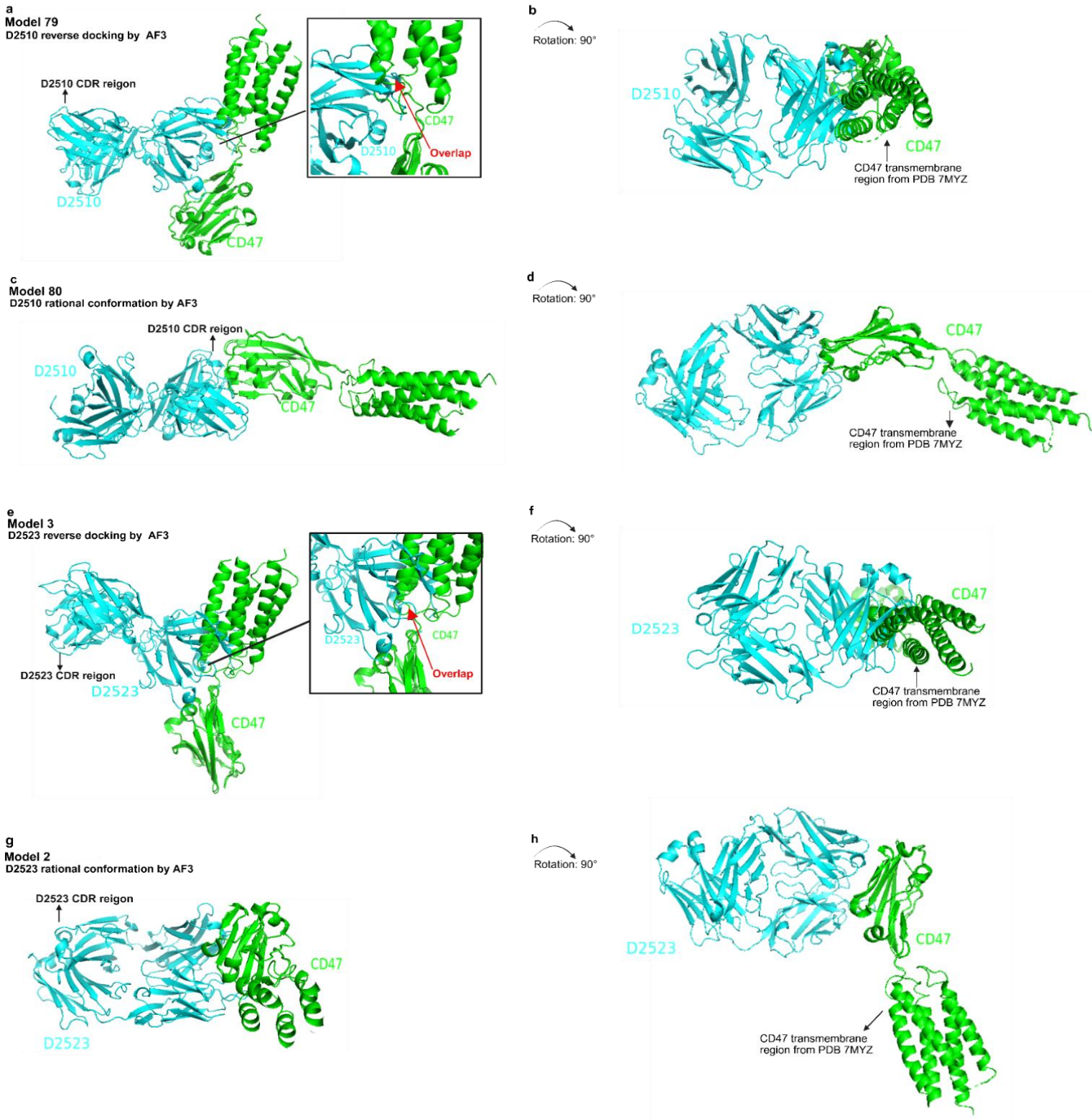

**a** Frontal view of the D2510-CD47 reverse docking conformation, where it can be seen that the non-complementarity determining region (CDR) region of D2510 is wrongly perceived as a binding region and binds to the part of the transmembrane region that connects to CD47 (ipTM=0.7, pTM=0.76) with detail of the incorrectly docked region in Model 79, where D2510 can be seen to Overlap with the CD47 transmembrane structure. **b** Rotation of panel a by 90° downwards view. **c** Frontal view of the D2510-CD47 Rational conformation, with the CDR region of D2510 docked at a reasonable angle to CD47 (ipTM=0.6, pTM=0.69). **d** Rotation of panel c by 90° downwards view. **e** Frontal view of the D2523-CD47 Reverse docking conformation, where it can be seen that the non-CDR region of D2523 is wrongly perceived as a binding region and binds to the part of the transmembrane region that connects to CD47 (ipTM=0.67, pTM=0.73) with detail of the incorrectly docked region in Model 79, where D2510 can be seen to Overlap with the CD47 transmembrane structure. **f** Rotation of panel e by 90° downwards view. **g** Frontal view of the D2523-CD47 Rational conformation, with the CDR region of D2510 docked at a reasonable angle to CD47 (ipTM=0.61, pTM=0.69). **h** Rotation of panel j by 90° downwards view. (CDR) is a specific region in an antibody molecule responsible for recognising and binding antigen.

## Figure 4. AF3 predictions for the D2510-CD47 (Full) and D2523-CD47 (Full)

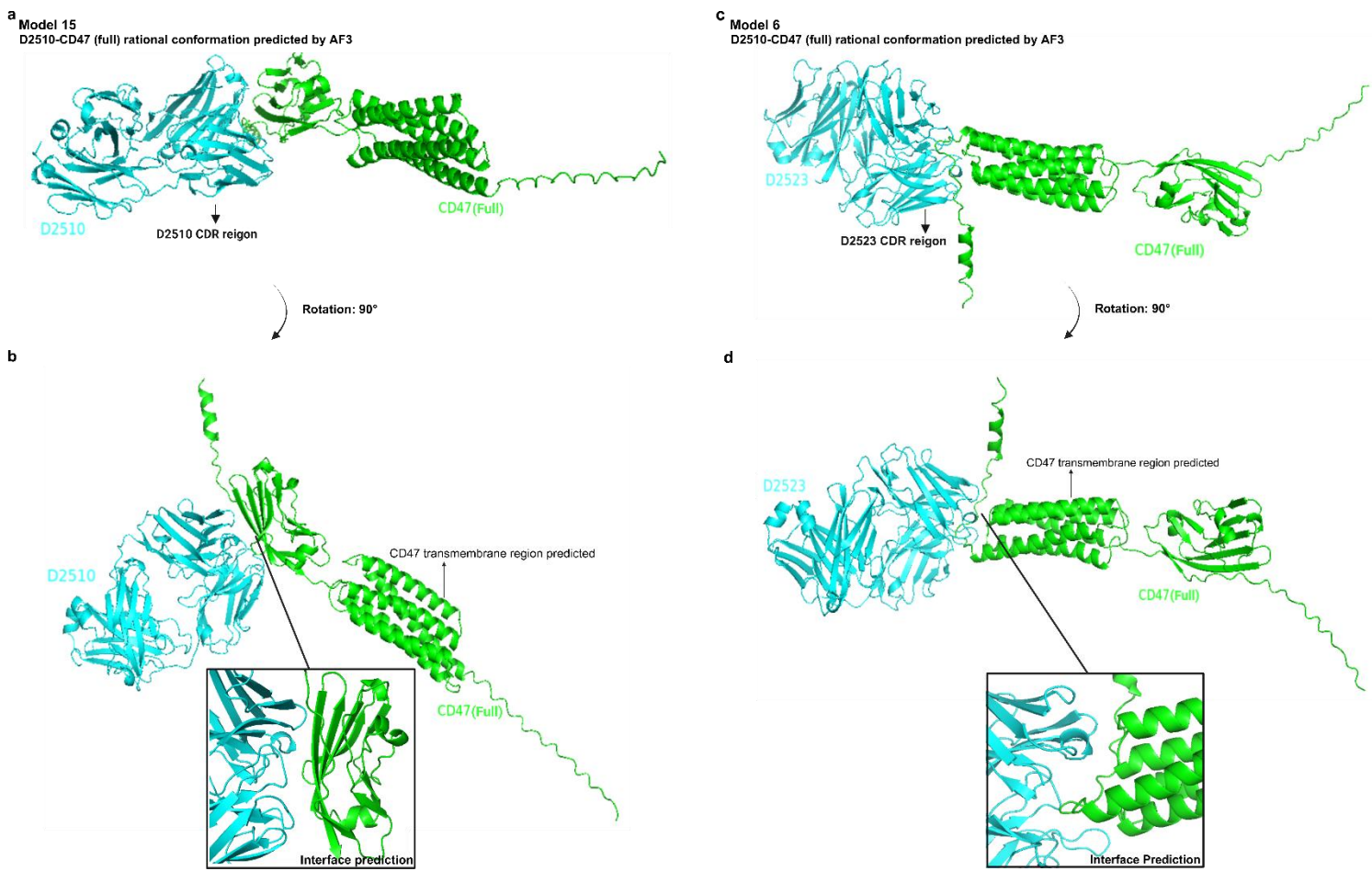

**a** Frontal view of D2510-CD47 (Full) and prediction of the 'Disorder structure' appearing in error (ipTM=0.43, pTM=0.56). **b** View of D2510-CD47 (Full) rotated down 90°, with interface details, showing reasonable side-to-side docking. **c** Frontal view of D2523-CD47 (Full) and 'Disorder structure' predicting the occurrence of errors, D2523 incorrectly docked CD47 transmembrane region (ipTM=0.47, pTM=0.58). **d** View of D2523-CD47 (Full) rotated down 180°, and interface details D2523 CDR, showing reasonable lateral docking with interface details D2523 CDR region irrationally interacts with the CD47 transmembrane region. CDR is a specific region in an antibody molecule responsible for recognising and binding antigen. Created in BioRender. Yiyang, X. (2025) https://BioRender.com/undefined

## Supporting Information

Additional experimental details, materials, and methods, including Molecular Model Files of All Subjects and Supplementary Materials. Molecular Model Files of All Subjects are sorted by three experiments with file names of Experiment 1-3 and additional Molecular Modelling file. Experiment 1-3 files contain all the PDB or CIF files of all subjects we used to conduct formal analysis in this study with corresponding data and description presented above. Molecular Modelling file contains four antibodies and CD47 antigen we used as inputs for subsequent calculations.

## Acknowledgments and Disclosure of Funding

We would like to thank the Antibody Engineering Laboratory of China Pharmaceutical University (CPU) for providing the necessary antibody sequences, wet-lab results and fundings. This is the first research mainly conducted and written by us (first co-authors Yiyang and Ziyou), as the beginning of our scientific career. Therefore, it is of great significance to us. We appreciate specifically Professor Juan Zhang and colleagues of the laboratory for advisements. At the same time, we admire School of Pharmacy and School of Science in CPU for conducting operational training on Discovery Studio and Schrodinger. Hefei Kejing Biotech Co., Ltd. provided technical support for AFM relevant structure predictions. Finally, we want to thank Doctor Dart Alwyn at UCL Cancer Institute for providing scientific guidance.

## Data and Software Availability

Data supporting the findings of this study are available in Supporting Information including Molecular Model Files of All Subjects and Supplementary Materials. All molecular structures or sequences of non-lab antibodies and antigen CD47 were extracted from online databases, including PDB: https://www.rcsb.org/ and Uniprot: https://www.uniprot.org/. All non-commercial experiments were conducted by online servers, including HDOCK server: http://hdock.phys.hust.edu.cn/, AF3 server: https://alphafoldserver.com/ and Swiss-Model server: https://swissmodel.expasy.org/.

## Author Contributions

**Yiyang Xu**: Conceptualization; formal analysis; project administration; resources; investigation; methodology; software; validation; data curation; visualization; writing – original draft; writing – review and editing. **Ziyou Shen**: Formal analysis; investigation; methodology; software; validation; data curation; visualization; writing – original draft; writing – review and editing. **Yanqing Lv**: Resources; validation; training; writing – review and editing. **Shutong Tan**: Resources; software; training. **Chun Sun:** Resources; training. **Juan Zhang:** Resources; funding acquisition; project administration; supervision; writing – review and editing.

## Conflict of Interest

The authors declare no competing financial interests.